\documentclass[12pt]{article}
\usepackage{epsfig}

\usepackage{amssymb}
\usepackage{amsmath}
\usepackage{amsfonts}

\usepackage{amsthm}

\theoremstyle{theorem}

\theoremstyle{definition}

\def\bp{\begin{proof}}
\def\ep{\end{proof}}

\def\be{\begin{equation}}
\def\ee{\end{equation}}
\def\ba{\begin{array}{c}}
\def\ea{\end{array}}

\def\ben{\[}
\def\een{\]}

\newcommand{\bea}{\begin{eqnarray}}
\newcommand{\eea}{\end{eqnarray}}

\newcommand{\bbr}{\br\!\br}

\newcommand{\kt}{\rangle}
\newcommand{\br}{\langle}
\begin{document}

\titlepage

\vspace{.35cm}

 \begin{center}{\Large \bf

The cryptohermitian smeared-coordinate
representation of wave functions

  }\end{center}

\vspace{10mm}

 \begin{center}

 {\bf Miloslav Znojil}

 \vspace{3mm}
Nuclear Physics Institute ASCR,

250 68 \v{R}e\v{z}, Czech Republic

{e-mail: znojil@ujf.cas.cz}

\vspace{3mm}


\end{center}

\vspace{5mm}


\section*{Abstract}

In discrete-coordinate quantum models the kinematical observable of
position  need not necessarily be chosen local (i.e., diagonal). Its
smearing is selected in the nearest-neighbor form of a real
asymmetric (i.e., cryptohermitian) tridiagonal matrix $\hat{Q}$. Via
Gauss-Hermite illustrative example we show how such an option
restricts the class of  admissible dynamical observables (sampled
here just by the Hamiltonian).


\section{Introduction \label{I} }

Virtually any textbook on quantum mechanics pays attention to the
point particle moving along straight line. Tacitly \cite{Hoo}, it is
assumed that the argument $x$ of its wave function $\psi(x) \in
L^2(\mathbb{R})$ coincides with an eigenvalue of an operator of the
(by assumption, observable) spatial position. The wave functions
$\psi(x)$ carry the standard probability-density interpretation and
describe the system in the so called $x-$representation.

For Hamiltonians $H=p^2+V(x)$ in which the external potential is
local, such an approach is natural. Due to the second-order
differential-operator form of $H$, one easily determines the
energies. The calculations may be further facilitated by the
transition to approximations. Typically \cite{discrsqw}, the
equidistant $N-$point Runge-Kutta discretization
 \ben
 x \in \mathbb{R} \ \ \longrightarrow \ \ x_j= x_j^{(RK)}=x_0^{(RK)}
 +h\cdot j\,,\ \ \ \ \ h >0\,,\ \ \
  {j=0,1,\ldots,N-1}
 \label{discrk}
 \een
converts operator $H$ into a real and symmetric tridiagonal $N$ by
$N$ matrix.

For non-local external potentials the concept of the coordinate
loses its guiding role of a link between the quantum and classical
pictures while the kinetic energy operator $p^2$ may still be
simplified via Fourier transformation ${\cal F}:\,\psi(x)\, \to\,
\tilde{\psi}(p)$. Thus, unless one has to deal with some other
observables defined as functions of $x$, the computing costs of the
mapping ${\cal F}$ remain acceptable.

The feasibility of the calculations worsens for the more complicated
Hamiltonians $H \neq p^2+V$ so that the role of the quality of the
approximations increases. In particular, the Runge-Kutta grid points
$x_j= x_j^{(RK)}$ may prove far from optimal in practice
\cite{amend}. Many non-Runge-Kutta discretizations are being
proposed and used in physics and quantum chemistry, therefore
\cite{Actonbe}.

\begin{table}[h]
\caption{The sample of the closed-form grid points $x_j^{(HP)}$ at
$N \leq 5$} \label{tpexp4}

\vspace{2mm}

\centering
\begin{tabular}{||c||cccccc||}
 \hline \hline
 $N$&$q_0$&$q_1$&$q_2$&$q_3$&$q_4$&\ldots\\
 \hline \hline
 1&0&&&&&\\
 2&$-{}{\sqrt {2}}{}$&${}{\sqrt {2}}{}$&&&&\\
 3&$-{}{\sqrt {6}}{}$&0&${}{\sqrt {6}}{}$&&&\\
 4&$-{}{\sqrt {6+2\,\sqrt {6}}}{}$&$-{}{\sqrt {6-2\,\sqrt {6}}}{}$&
 ${}{\sqrt {6-2\,\sqrt {6}}}{}$&${}{\sqrt {6+2\,\sqrt {6}}}{}$&&\\
 5&$-{}{\sqrt {10+2\,\sqrt {10}}}{}$&$-{}{\sqrt {10-2\,\sqrt {10}}}{}$&0
 &${}{\sqrt {10-2\,\sqrt {10}}}{}$&${}{\sqrt {10+2\,\sqrt {10}}}{}$&\\
 \vdots&\ldots&&&&&\\
 \hline \hline
\end{tabular}

\end{table}

In our forthcoming methodical considerations one of the most typical
samples of the non-Runge-Kutta one-dimensional grid points will be
selected in the form of the $N-$plet of the zeros of the $N-$th
Hermite polynomial \cite{AS},
 \be
 x \in \mathbb{R} \ \ \longrightarrow \ \ x_j= x_j^{(HP)}\,,
 \ \ \ \ \ \
 {j=0.1,\ldots,N-1}\,.
 \label{discrhp}
 \ee
In a way explained in Appendix A we felt puzzled by the conflict
between the amendment of the numerical efficiency gained by the
transition from $x_j^{(RK)}$ to $x_j^{(HP)}$ (cf.
Refs.~\cite{AS,uNIST} for details) and the loss of the closed
formulae for $x_j^{(HP)}$ (with a few low$-N$ exceptions sampled in
Table~\ref{tpexp4}). A softening of the latter disadvantage (i.e.,
of the purely numerical character of the amended values $x_j^{(HP)}$
at higher $N$) will be proposed here, therefore.

In section \ref{II}, first of all, the  numerical values of
$x_j^{(HP)}$ will be reinterpreted as the implicitly defined
eigenvalues of a suitable tridiagonal matrix $\hat{Q}^{(N)}$ with
elementary matrix elements. It will be argued there that one of
versions of such a next-to-diagonal form of $\hat{Q}^{(N)}$ is in
fact often used during the evaluation of values $x_j^{(HP)}$ in
numerical practice.

In section \ref{paragrafy} (complemented also by Appendix B) our
specific choice of manifestly non-Hermitian $N$ by $N$ matrices
$\hat{Q}^{(N)}$ will be advocated. Their extreme simplicity will be
identified there with the main criterion of applicability of the
increasingly popular use of representations of quantum observables
(sampled here by the position and energy) in the so called
cryptohermitian picture.

In section \ref{III}, the main formal aspects and consequences of
the use of the grid $x_j^{(HP)}$ in such an implicit representation
will be explained, in detail, via the first nontrivial $N=4$ model.
The abstract picture of kinematics (represented by the matrix
$\hat{Q}^{(N)}$) and dynamics (represented by the related admissible
families of Hamiltonians $H^{(N)}$ and/or of the so called metric
operators $\Theta^{(N)}$) will be given there the concrete forms in
which the construction of matrices $\Theta^{(N)}$ in the simplest
(viz., tridiagonal and pentadiagonal) sparse-matrix forms will be
paid particular attention.

In the final section \ref{V} the resulting change of perspective
transferring emphasis from the simplicity of quantum kinematics to a
potentially better balance between the simplicity of kinematics and
dynamics will be summarized.

\section{Formalism: Kinematics  \label{II}}

\subsection{An implicit definition of grid points $x_j^{(HP)}$ \label{IIa}}

Let us consider the classical orthogonal Hermite polynomials
 \be
 H_0(x)= 1\,,\ \ \ H_1(x)= 2\,x
 \,,\ \ \ H_2(x)=
 4\,{x}^{2}-2
 \,,\ \ \ H_3(x)=
 8\,{x}^{3}-12\,x
 \,,\ldots\,
 \label{same}
 \ee
and make our present grid points $x_{j}^{(HP)}$ unique by defining
them as the roots of equation
 \be
 H_N\left (\frac{1}{2}\,x_j^{(HP)}\right )=0\,,\ \ \ \
 j = 0, 1, \ldots\,, N-1\,.
 \label{sec}
 \ee
Naturally, such a definition of the grid points is purely numerical,
at the larger lattice-sizes $N$ at least. At the finite $N$, there
is in fact no practical necessity of insisting on the
multiplicative-operator nature (i.e., on the diagonal matrix form
and representation) of the position-operator
$\hat{Q}=\hat{Q}^{(N)}$. Thus, once we re-read our implicit
lattice-definition (\ref{sec}) as a vanishing-condition for the
determinant of recurrences for Hermite polynomials we may
immediately replace the traditional and, in essence, purely
numerical diagonal representation of the operator $\hat{Q}^{(N)}$ by
its following, well known tridiagonal-matrix alternative
  \be
 \hat{Q}^{(N)}=\left[ \begin {array}{cccccc}
 0&1&0&0&\ldots&0
\\
\noalign{\medskip}2&0& 1&0&\ddots&\vdots
\\
\noalign{\medskip}0&4&0& \ddots&\ddots&0
\\
\noalign{\medskip}0&0&6&\ddots& 1&0
\\
\noalign{\medskip}\vdots&\ddots&\ddots&\ddots&0& 1
\\
\noalign{\medskip}0&\ldots&0&0&{2N-2}&0
\end {array} \right]\,.
 \label{hamilH}
 \ee
The proof of the coincidence of the eigenvalues of this matrix with
the roots of Eq.~(\ref{sec}) is easy - it suffices to recall the
recurrences for Hermite polynomials \cite{Maple}. Subsequently, we
may even specify the related eigenvectors of matrix (\ref{hamilH}),
i.e., the smeared position eigenstates in closed form,
 \be
 |\psi_n\kt
 =
 \left (
  \ba
 H_0(x_n/2)\\
 H_1(x_n/2)\\
 \vdots\\
 H_{N-1}(x_n/2)
 \ea
 \right )\,,\ \ \ \ \  n = 0, 1, \ldots\,, N-1\,.
 \label{inputH}
 \ee
In such an overall setting one has to circumvent several conceptual
as well as purely technical obstacles. Some of them will be
discussed in what follows.

\subsection{A trivial Hermitization of  $\hat{Q}^{(N)}$ \label{IIIb}}

As long as we have  $\hat{Q}^{(N)}\neq \left [\hat{Q}^{(N)}\right
]^\dagger$ we must abandon the ``friendly" real vector space ${\cal
H}^{(F)}=\mathbb{R}^N$ as unphysical. At the same time we may
introduce a diagonal $N$ by $N$ matrix $\Omega^{(N)}_0$ with
positive matrix elements along its diagonal, $ \left
[\Omega^{(N)}_0\right
 ]_{mn} =
\omega_n\delta_{mn}$. This enables us to define matrix
 \be
 \mathfrak{q}^{(N)}_0=\Omega^{(N)}_0\hat{Q} \left [\Omega^{(N)}_0\right
 ]^{-1}\,
 \label{gri}
 \ee
possessing the same eigenvalues as $\hat{Q}$. As long as the
original position operator $\hat{Q}$ itself is a tridiagonal matrix,
we are allowed to require that our grid-points-preserving similarity
transformation (\ref{gri}) has the Hermitization property,
 \ben
 \mathfrak{q}^{(N)}_0 =
\left [ \mathfrak{q}^{(N)}_0
 \right ]^\dagger\,.
  \een
We may put $\omega_n=c/\sqrt{(2n)!!}$ at any $c \neq 0$ and $n=0, 1,
\ldots, N-1$ yielding the closed and well known formula
  \be
 \mathfrak{q}^{(N)}_0=\left[ \begin {array}{cccccc}
 0&\sqrt{2}&0&0&\ldots&0
\\
\noalign{\medskip}\sqrt{2}&0& \sqrt{4}&0&\ddots&\vdots
\\
\noalign{\medskip}0&\sqrt{4}&0& \ddots&\ddots&0
\\
\noalign{\medskip}0&0&\sqrt{6}&\ddots& \sqrt{2N-4}&0
\\
\noalign{\medskip}\vdots&\ddots&\ddots&\ddots&0& \sqrt{2N-2}
\\
\noalign{\medskip}0&\ldots&0&0&\sqrt{2N-2}&0
\end {array} \right]\,.
 \label{hamilHui}
 \ee
Formally speaking \cite{SIGMA} we may now treat the new matrix
$\mathfrak{q}^{(N)}_0 $ as acting in an idealized, ``physical"
Hilbert space ${\cal H}^{(P)}_0$ which remains isomorphic to
$\mathbb{R}^N$.

\section{Formalism: Dynamics \label{paragrafy} }

\subsection{Hamiltonians \label{paragra} }

The standard textbook picture of quantum dynamics may be perceived
as living in Hilbert space ${\cal H}^{(P)}_0$. In this space the
time evolution may be assumed generated by an arbitrary ``effective"
Hermitian Hamiltonian matrix $\mathfrak{h}^{(N)}_{0,e}=\left
[\mathfrak{h}^{(N)}_{0,e}\right ]^\dagger$. Its isospectral backward
map
 \be
 H^{(N)}_{0,e}=\left [\Omega^{(N)}_0\right ]^{-1}
 \mathfrak{h}^{(N)}_{0,e}\Omega^{(N)}_0
 \label{grise}
 \ee
will then define the non-Hermitian Hamiltonian acting in the old
``friendly" space ${\cal H}^{(F)}$. The original Hermiticity of
$\mathfrak{h}^{(N)}_{0,e}$ is strictly equivalent to the
Dieudonn\'{e}'s \cite{Dieudonne} quasi-Hermiticity relation
rewritten in the double-dagger-superscripted notation and in terms
of the abbreviation $\Omega^\dagger_0\Omega_0:=\Theta_0$,
 \be
 H^{(N)}_{0,e}=\left [H^{(N)}_{0,e}\right ]^\ddagger
 :=\left [\Theta^{(N)}_0\right ]^{-1}
 \left [H^{(N)}_{0,e}\right ]^\dagger
  \Theta^{(N)}_0\,.
  \label{newde}
 \ee
Under the implicit methodical assumption that the two operators of
observables $ \mathfrak{q}^{(N)}_0 $ (= position) and $
\mathfrak{h}^{(N)}_{0,e} $ (= Hamiltonian) in ${\cal H}^{(P)}_0$ are
``prohibitively complicated", the determination of at least some of
the properties of the quantum system in question may still prove
simpler in the unphysical, auxiliary Hilbert space ${\cal H}^{(F)}$
where the isospectral operators of observables $ \hat{Q}^{(N)} $ (=
position) and $ H^{(N)}_{0,e} $ (= Hamiltonian) appear manifestly
non-Hermitian. Thus, their {\em direct} use in computations (e.g.,
of their spectra) may happen to remain well motivated \cite{Geyer}.

\subsection{The introduction of the third Hilbert space}

Let us now assume that at any fixed integer $\alpha \geq 1$ and in
all of the text of paragraphs \ref{IIIb} and \ref{paragra} one
replaces the {\em diagonal} (i.e., trivial) matrix $\Omega^{(N)}_0$
by a fully general matrix $\Omega^{(N)}_\alpha$ (to be called a
Dyson's map \cite{Geyer}) such that the product $\Theta_\alpha
=\Omega^\dagger_\alpha\Omega_\alpha$ (i.e., a certain less trivial
metric) remains sparse and strictly $(2\alpha+1)-$diagonal. Under
this assumption the {\em kinematics} of our schematic quantum system
may still be represented by {\em the same} position matrix $
\hat{Q}^{(N)} $ defined in ${\cal H}^{(F)}$. Naturally, the same
specification of kinematics (i.e., of the grid points) also
reappears, in the {\em new} physical Hilbert space ${\cal
H}^{(P)}_\alpha$, as carried by the {\em new}, manifestly Hermitian
matrix $\mathfrak{q}^{(N)}_\alpha$.

Naturally, the transition from $\alpha=0$ to $\alpha > 0$ will allow
us to {\em change the dynamics} in general. This is the key idea of
our present paper. Thus, the choice of any $\alpha-$dependent
{dynamical} input information represented, say, by some ``effective"
Hamiltonian $\mathfrak{h}^{(N)}_{\alpha,e}$ remains fully at our
disposal. In our present exemplification of the theory this
Hamiltonian will be chosen as a matrix which is Hermitian in ${\cal
H}^{(P)}$ and which determines the observable properties of our
hypothetical, $\alpha-$parametrized quantum system.

The related generalized, $\alpha-$parametrized version of the
non-diagonal Dyson's map $\Omega^{(N)}_\alpha$ will be again, in
general, non-unitary, implying that $\Theta_\alpha\neq I$. This
means that in a close parallel to the preceding two paragraphs where
we used $\alpha=0$, we shall be forced to declare the ``friendly"
Hilbert space ${\cal H}^{(F)}_0$ ``false" and {\em unphysical}.

In such a case, in a way outlined in \cite{SIGMA}, it proves useful
to introduce the third, ``standardized" Hilbert space ${\cal
H}^{(S)}_\alpha$. By definition, the latter space may {\em coincide}
with its ``friendly" partner ${\cal H}^{(F)}$ as a vector space of
kets $|\psi\kt$. At the same time it will certainly {\em differ}
from it by its $\Theta_\alpha-$dependent definition of the Hermitian
conjugation of operators.

Formally, we may merely replace the zero subscripts in
Eqs.~(\ref{gri}), (\ref{grise}) and (\ref{newde}) by their
generalized forms, $ _0 \longrightarrow _\alpha$, while calling the
positive definite operators $\Theta_\alpha=\Theta_\alpha^\dagger$
the ``Hilbert-space metric operators" \cite{Geyer}. Indeed, by
construction, the physics described in the third Hilbert space
${\cal H}^{(S)}_\alpha$ remains strictly the same as in ${\cal
H}^{(P)}_\alpha$. The unitary equivalence between Hilbert spaces
${\cal H}^{(S)}_\alpha$ and ${\cal H}^{(P)}_\alpha$ may be also
emphasized by the notation recommended in Ref. \cite{SIGMA} and
defining the linear functionals (a. k. a. the Dirac's bra-vectors
$\br \psi|^{(S)}$ or rather, in the compactified notation of Ref.
\cite{SIGMA}, ``brabravectors") in the less usual Hilbert space
${\cal H}^{(S)}_\alpha$ via the Hermitian-conjugation operation
which is $\alpha-$dependent and reads
 \be
 {\cal T}^{(S)}_\alpha: |\psi\kt \ \to \ \br \psi|^{(S)}
 :=\bbr \psi|=\br
 \psi|\Theta_\alpha\,.
 \label{nediraco}
 \ee
The values of the inner products $\bbr \psi|\phi\kt$ in the
``standardized" physical Hilbert space ${\cal H}^{(S)}$ may be also
evaluated inside the auxiliary Hilbert space ${\cal H}^{(F)}$ since
 \be
 \bbr \psi|\phi\kt \left |_{in \ {\cal
H}^{(S)}}\  \equiv \ \br \psi|\Theta|\phi\kt\right |_{in \ {\cal
H}^{(F)}}\,.
 \label{formulas}
 \ee
Thus, we never leave the standard quantum mechanics. The ${\cal
H}^{(S)}-$specifying metric must, of course, have all of the
required properties (i.e., basically, $\Theta=\Theta^\dagger>0$ in
the present scenario with $N < \infty$ \cite{Geyer}). In other
words, the requirement of the consistency of the theory may be
expressed, inside the auxiliary Hilbert space ${\cal H}^{(F)}$, as
the compatibility of our  matrices of observables (viz.,  position
$\hat{Q}= \hat{Q}^{(N)} $ and Hamiltonian $H= H^{(N)}_{\alpha,e} $)
with the Dieudonn\'{e}'s constraints
 \be
 \hat{Q}^\dagger\,\Theta=\hat{Q}\Theta \,
 \label{dieudo}
 \ee
 \be
 H^\dagger\,\Theta=H\Theta \,.
 \label{dieudobe}
 \ee
They represent just the necessary conditions of the (hidden)
Hermiticity of the respective operators of observables.

\section{An illustrative example with $N=4$ \label{III}}

\subsection{General non-diagonal isocoordinate mappings $\Omega^{(4)}$}

At one of the simplest nontrivial choices of $N=4$ the present
``smeared" (i.e., non-diagonal) operator of position reads
 $$
 \hat{Q}^{(4)}=
 \left[ \begin {array}{cccc} 0&1&0&0\\\noalign{\medskip}2&0&1&0
\\\noalign{\medskip}0&4&0&1\\\noalign{\medskip}0&0&6&0\end {array}
 \right].
 $$
The related exhaustive (i.e., four-parametric) solution $\Theta$ of
Eq.~(\ref{dieudo}) is then easily found to read
 \be
\Theta^{(4)}_{(k,{\mu},p,d)}=\left[ \begin {array}{cccc}
k&{\mu}&p&d\\\noalign{\medskip}{\mu}&1/2\,k+2\,p&1/2\,{\mu}+3\,d&1/2\,p\\
\noalign{\medskip}p&1/2\,{\mu}+3\,d&p+1/8\,k&1/2\,d+1/
8\,{\mu}\\\noalign{\medskip}d&1/2\,p&1/2\,d+1/8\,{\mu}&1/12\,p+1/48\,k
\end {array} \right]
\label{result}
 \ee
Once we select $k=1$ and ${\mu}=p=d=0$ we re-obtain the
above-discussed $\alpha=0$ metric $\Theta_0$ with elements $\left
[\Theta_0\right ]_{11} =1$, $\left [\Theta_0\right ]_{22} =1/2$,
$\left [\Theta_0\right ]_{33} =1/8$ and $\left [\Theta_0\right
]_{44} =1/48$. In an opposite extreme of the fully general
four-parametric metric (\ref{result}), one can find the complete set
of the underlying isocoordinate mappings $\Omega^{(4)}$ via the
factorization of $\Theta=\Omega^\dagger\Omega$.

The inspection of the second independent constraint~(\ref{dieudobe})
reveals that in the trivial diagonal-metric example with
$\Theta=\Theta_0^{(N)}$ an important subset of the admissible
Hamiltonians $H$ will be composed of arbitrary diagonal real
matrices. In such a setting one  might speak about the (discrete)
wave functions in energy-representation.

\subsection{Tridiagonal one-parametric metric}

At $k=1$ and $p=d=0$ the  $\alpha=1$ metric found from
Eqs.~(\ref{result}) or (\ref{dieudo}) reads
 \be
 \Theta_1^{(4)}=\Theta_1^{(4)}({\mu})=
  \left[ \begin {array}{cccc} 1&{\mu}&0&0\\\noalign{\medskip}{\mu}&1/2&1/2\,{\mu}&0
\\\noalign{\medskip}0&1/2\,{\mu}&1/8&1/8\,{\mu}\\\noalign{\medskip}0&0&1/8\,{\mu}&
1/48\end {array} \right]\,.
 \label{mua}
 \ee
It might be factorized, in particular, into products
$\Theta=\Omega^\dagger\Omega$ using the lower-two-diagonal ansatz
for the real factor $\Omega$ and evaluating its matrix elements in
the recurrent manner starting from the lower corner.

\begin{figure}[h]                     
\begin{center}                         
\epsfig{file=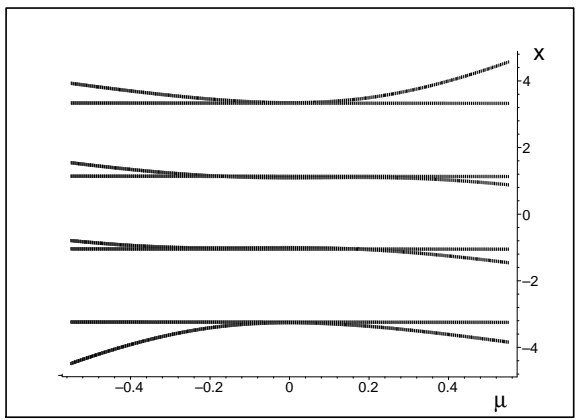,angle=270,width=0.5\textwidth}
\end{center}                         
\vspace{-2mm}\caption{The quality of approximation of the
$\mu-$independent spectrum of $\hat{Q}^{(4)}$ by the $\mu-$dependent
grid-point eigenvalues of the Hermitian position matrix
$\hat{q}^{(4)}(\mu)$ considered in the leading-order tridiagonal
small$-\mu$ approximation (\ref{eqyui}).
 \label{fi2}}
\end{figure}

Once we wish to avoid the recurrent factorizations (which do not
lead to any nice formulae at general $N$), a direct algebraic method
may be used whenever the off-diagonal part of the metric remains
small. For illustration let us notice that in such an approach one
reproduces the metric of Eq.~(\ref{mua}) by the approximate $N=4$
and $\alpha=1$ matrix
 $$
 \Omega({\mu})\approx \left[ \begin {array}{cccc}
 1-{{\mu}}^{2}&0&0&0\\
 \noalign{\medskip}{\mu}
\sqrt {2} \left( 1+2\,{{\mu}}^{2} \right) &\frac{1}{2}\,\sqrt {2}
\left(
1-2\,{{\mu}}^ {2} \right) &0&0\\
\noalign{\medskip}0&{\mu}\sqrt {2} \left( 1+3\,{{\mu}}^{2}
 \right) &\frac{1}{4}\,\sqrt {2} \left( 1-3\,{{\mu}}^{2} \right) &0
\\
\noalign{\medskip}0&0&\frac{1}{2}\,{\mu}\sqrt
{3}&\frac{1}{12}\,\sqrt {3}\end {array}
 \right]
 $$
which is exact up to the fourth-order uncertainty factor $1+{\cal
O}({\mu}^4)$. In the same perturbative spirit we may accept the
larger uncertainty $1+O \left( {{\mu}}^{2} \right )$ in the
approximate inverse
 $$
 \Omega^{-1}({\mu})\approx
 \left[ \begin {array}{cccc}
  1+{{\mu}}^{2}&0&0&0\\
  \noalign{\medskip}-2\,{\mu}
 \left( 1+{{\mu}}^{2} \right) &\sqrt {2} \left( 1+2\,{{\mu}}^{2} \right) &0&0
 \\
 \noalign{\medskip}0&-4\,{\mu}\sqrt {2} \left( 1+2\,{{\mu}}^{2} \right) &2\,
 \sqrt {2} \left( 1+3\,{{\mu}}^{2} \right)
 &0\\
 \noalign{\medskip}0&0&-12\,{\mu} \sqrt {2} \left( 1+3\,{{\mu}}^{2}
 \right) &4\,\sqrt {3}
  \end {array}
 \right]
 $$
and obtain the first-order correction to formula (\ref{hamilHui})
above,
 \be
 \mathfrak{q}^{(N)}_1({\mu}) \approx \left[ \begin {array}{cccc}
 -2\,{\mu}  &\sqrt {2}
 &0&0\\
 \noalign{\medskip}
 \sqrt {2}
 &-2\,{\mu}&
 2
 &0\\
 \noalign{\medskip}0&2&-2\,{\mu}&\sqrt {2}\sqrt {3}
\\
\noalign{\medskip}0&0&\sqrt {2}\sqrt {3}&6\,{\mu}
\end {array} \right]\,.
\label{eqyui}
 \ee
In this precision the Hermitized position matrix is tridiagonal. The
grid-point eigenvalues evaluated from the approximate position
matrix (\ref{eqyui}) coincide with the exact values at the small
$\mu$ (cf. Figure \ref{fi2}).

\begin{figure}[h]                     
\begin{center}                         
\epsfig{file=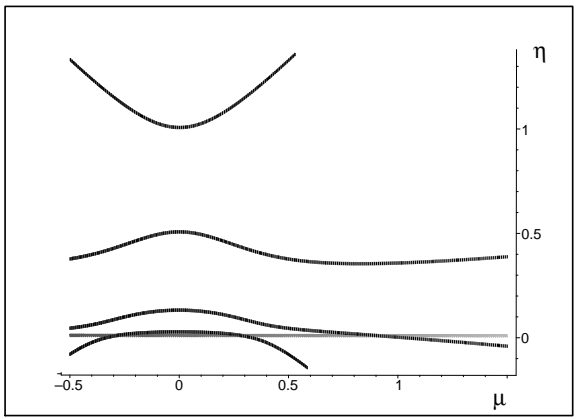,angle=270,width=0.5\textwidth}
\end{center}                         
\vspace{-2mm}\caption{The ${\mu}-$dependence of the quadruplet of
the eigenvalues of the tridiagonal candidate (\ref{mua}) for the
metric.
 \label{fi0}}
\end{figure}

The consistency of the perturbation calculations with respect to the
small ${\mu}$ may be reconfirmed by the inspection of Figure
\ref{fi0} in which one sees that the smallest eigenvalue of the
exact one-parametric $\alpha=1$ (i.e., tridiagonal) metric
(\ref{mua}) ceases to be positive at the not too large values of
$|{\mu}| \approx 0.3$. The exact formula
 \ben
  \Theta_1^{(4)}({\mu})> 0 \ \ {\rm iff} \ \ {\mu} \in  \ \left (-{\mu}_0,{\mu}_0 \right)\,,
  \een
  \be
  {\mu}_0=
  1/\sqrt{6+2\,\sqrt{6}}
  \approx 0.3029054464
  \label{bounda}
  \ee
is also available due to the exact solvability of the model at
$N=4$.

\subsection{Two-parametric pentadiagonal metric}

Our above commentary accompanying Figure \ref{fi0} means that at the
larger $\mu$ the requirement of the positivity of the metric can
only be satisfied at $\alpha \geq 2$, i.e., in a less trivial
physical domain ${\cal D}$ of the available variable parameters $k$,
${\mu}$, $p$ and $d$ entering the fully general formula of
Eq.~(\ref{mua}) for the metric.

\begin{figure}[h]                     
\begin{center}                         
\epsfig{file=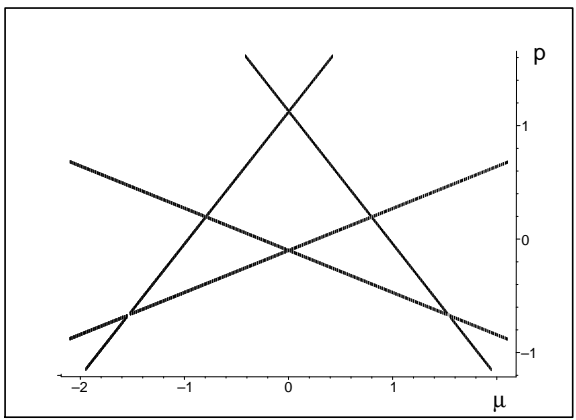,angle=270,width=0.5\textwidth}
\end{center}                         
\vspace{-2mm}\caption{The boundaries of the domain of positivity of
the pentadiagonal candidate for the metric.
 \label{fi1}}
\end{figure}

The simplicity of our $N=4$ model enables us to make this argument
more quantitative. The explicit $\alpha=2$ choice of the two
variable parameters $\mu$ and $p$ in the metric
 $$
 \Theta_2^{(4)}({\mu},p)= \left[ \begin {array}{cccc}
 1&{\mu}&p&0\\\noalign{\medskip}{\mu}&1/2+2\,
 p&1/2\,{\mu}&1/2\,p\\\noalign{\medskip}p&1/2\,{\mu}&p+1/8&1/8\,{\mu}
\\\noalign{\medskip}0&1/2\,p&1/8\,{\mu}&1/12\,p+1/48\end {array} \right]
 $$
can be used for this purpose. In a preliminary graphical analysis we
reveal (cf. Figure \ref{fi1}) that the two-dimensional version of
the physical domain ${\cal D}$ has the piecewise linear boundaries.
A more rigorous explanation can be also provided since the boundary
lines may be identified with the loci of zeros of the secular
polynomial $S({\mu},p,\eta) = \det (\Theta_2^{(4)}({\mu},p)-\eta)$
at $\eta=0$. Once we evaluate
 $$
 S({\mu},p,0) = \det \Theta_2^{(4)}({\mu},p)=
  $$
  $$
  ={\frac {1}{768}}-1/8\,{p}^{2}{{\mu}}^{2}
 -1/6\,{p}^{3}+1/16\,{p}^{2}+1/12\,
 {p}^{4}+1/48\,p-{\frac {1}{64}}\,{{\mu}}^{2}+{\frac {1}{64}}\,{{\mu}}^{4}
 $$
we reveal that the polynomial $S({\mu},p,0)$ factorizes. This
explains the existence of the two left-right symmetric pairs of the
straight nodal lines which cross at ${\mu}=0$. The upper one is
prescribed by the equation
 $$
 {\mu}_{upper} \sim \pm (p-p_{upper})
 = \pm \left (p-{\frac {1}{2\,\sqrt {6}-4}}\right )
  $$
(in Fig.~\ref{fi1}, this is the upper pair crossing at the positive,
maximal admissible $p_{upper}=1.112372435$). Similarly, the lower
pair of lines given by the equation
$$
 {\mu}_{lower} \sim \pm (p-p_{lower})
 = \pm \left (p+{\frac {1}{2\,\sqrt {6}+4}}\right )
  $$
crosses at the minimal,  negative $p_{lower}=-0.1123724357$.

Along the line $p=0$, Fig.~\ref{fi1} correctly reproduces  the
position of the two boundaries (\ref{bounda}) of the interval  of
the positivity of all of the eigenvalues of $\Theta_1^{(4)}({\mu})$
as displayed in Fig.~\ref{fi0}. The same picture also explains why
this interval of positivity of the metric grows with $p>0$ up to
$p_{maximal} = 1$.

\section{Summary \label{V}}

In accord with all textbooks on quantum mechanics the set of the
operators of observables (sampled here by the position
$\mathfrak{q}$ and energy $\mathfrak{h}$) must be kept Hermitian in
a certain physical Hilbert space of states (here, in ${\cal
H}^{(P)}$). In such a setting the starting idea of our present paper
was that such a Hilbert space may be replaced by any unitarily
equivalent alternative (in our notation, by ${\cal H}^{(S)}$).

The majority of the existing applications of such an idea is built
upon the unitary, Fourier-like correspondence ${\cal F}$ between bra
or ket elements of the two spaces. In our present paper we rather
made use of an indirect, manifestly non-unitary correspondence
$\Omega$ involving just the ket vectors of the two spaces
\cite{SIGMA}. As far as we know (cf. also several additional
comments collected in Appendix~B), the first use of such a
correspondence (called Dyson map) found its applications in nuclear
physics \cite{Geyer}.

Our present form of Hilbert spaces ${\cal H}^{(P)}$ or ${\cal
H}^{(S)}$ has been chosen finite-dimensional, generated via a
systematic $N-$point discretization of the real line of coordinates.
The underlying idea was that in such a setting the efficiency of the
calculations need not necessarily be optimal when we diagonalize, by
brute force, all of the matrices in question (i.e., here, the pair
of matrices $\mathfrak{q}$ and $\mathfrak{h}$). We imagined that
even the Hermiticity of the matrices of observables need not be a
condition {\it sine qua non}. We argued that the non-diagonal forms
of these matrices, whenever sufficiently compact, would offer a way
towards a minimization of the time needed for the numerical
evaluations of phenomenological predictions.

For illustrative purposes we started from the choice of the
``kinematic" grid-point eigenvalues $x_j^{(HP)}$ of a matrix
$\mathfrak{q}$. We imagined that the latter matrix may be further
simplified via a nonunitary Fourier-like transformation
$\Omega^{-1}$,
 \be
 {\cal H}^{(P)}\ \longrightarrow\   {\cal H}^{(F)}\,.
 \label{prvni}
 \ee
In the latter, naively chosen Hilbert space the isospectral analogue
$\hat{Q}$ of the position appears non-Hermitian but the remedy is
easy. In the subsequent ``Hermitization" step
 \be
 {\cal H}^{(F)}\ \longrightarrow\   {\cal H}^{(S)}\,
 \label{druha}
 \ee
one simply endows the old space with a new, nontrivial metric
$\Theta\neq I$, i.e., with an alternative, amended inner product.

Our present attention has been paid to the definition of the third
space ${\cal H}^{(S)}$. We showed that and how the trick
(\ref{prvni}) plus (\ref{druha}) opens a new paradigm in the
constructive analysis of quantum dynamics by making the pullback
$\hat{Q}\ \longrightarrow\ \mathfrak{q}$ internally consistent. We
showed that even though our pre-selected operators $\mathfrak{q}$ or
$\hat{Q}$ of the coordinate were not diagonal, their respective
Hermiticity inside ${\cal H}^{(P)}$ and ${\cal H}^{(S)}$ has still
been, {\em constructively}, guaranteed.

On such a purely kinematical background the alternative dynamics
were assumed given by the variable class of energy observables
$\mathfrak{h}$. We explained how these Hamiltonian matrices may be
related to their cryptohermitian avatars $H$, and how their
eigenfunctions might be assigned the standard probabilistic
interpretation. Thus, in contrast to the existing literature as
reviewed, say, in \cite{Carl} and starting from the wave functions
in $x-$representation, we did not postulate any Hamiltonian in
advance. We merely fixed its eligible class compatible with the
pre-selected, smeared-coordinate representation of the wave
functions.

Our main emphasis has been put on a {\em minimal} smearing of the
position. For the entirely pragmatic purposes we selected the
$N-$plet of grid points $x_j$ adapted to the optimal precision of
the (discrete, approximate) integrations. Our matrices
$\hat{Q}^{(N)}$ have been chosen tridiagonal, representing just one
of the most elementary lattice grids given as the Hermite-polynomial
zeros. We expect that such a strategy will optimize the precision of
approximants, especially when evaluating or prescribing the norms
and/or inner products of wave functions.

Summarizing, we proposed here the kinematical choice of the
cryptohermitian position $\hat{Q}^{(N)}$ accompanied by the
dynamics-determining specification of one of the eligible metrics
$\Theta_\alpha$. In such a framework we indicated the possibility of
varying the dynamics and/or of moving towards a systematic and
exhaustive construction and classification of all of the
alternative, $\alpha-$numbered physical Hilbert spaces, be it ${\cal
H}^{(P)}_\alpha$ and/or its unitary equivalent, less usual version
${\cal H}^{(S)}_\alpha$ with nontrivial metrics given, in our
illustrative example, in closed form.

\section*{Appendix A: A comment on choices of grid points }

One of the roots of the popularity of the representation of the
quantum-state ket-vectors $|\psi\kt$ in their wave-function (alias
``coordinate-representation") form $\psi(x)\in L^2(\mathbb{R},\mu)$
may be seen in the user-friendliness of the formula giving their
norm,
 \be
 \|\psi\|:=\sqrt{\int_{-\infty}^\infty\,\psi^*(x)\psi(x){\omega}(x)
 dx}\,.
 \label{norms}
 \ee
In a way similar to the definition of the Hilbert-space inner
product
 \ben
 \br \psi|\phi \kt:=\int_{-\infty}^\infty\,\psi^*(x)\phi(x){\omega}(x)dx\,
 \een
these integrals contain a weighting function ${\omega}(x)$ which is
most often set equal to a constant or a power of $x$ or a Gaussian.

The necessity of the evaluation of these integrals is ubiquitous.
Its precision represents one of the limiting factors of the
practical applicability of the theory. In particular, various
empirical tests are very often passed by the so called Gauss-Hermite
quadrature \cite{AS}
 \be
 \int_{-\infty}^\infty\,e^{-x^2}\,f(x) \,dx\ = \sum_{j=0}^{N-1}\,
 w_j\,f(x_j) + error[N,f^{(2N)}(\xi)]\,.
 \ee
In such a Gaussian-weighted approximative formula the size of the
error term is very often found acceptably small over the relevant
class of the integrands. Its grid-point quantities $x_j$ are
identified with the the roots of the well known Hermite polynomials
(this identification is sampled by our present Eq.~(\ref{sec})
above). The necessary weight factors are also obtainable in compact
form, $w_j=w_j^{(N)}=C(N) /[H(N-1,x_j)]^2$. The more detailed
information (i.e., the proportionality factor $C(N)$, etc) may be
found not only in standard  monographs \cite{Acton} but also,
directly and free of charge, via web~\cite{beNIST,NIST}.

\section*{Appendix B: A comment on the concept of cryptohermiticity}

In the Dieudonn\'{e}'s paper \cite{Dieudonne} operators $\hat{Q}$
satisfying relation (\ref{dieudo}) at a suitable positive and
self-adjoint ``metric" $\Theta$ were studied and given the name of
``quasi-Hermitian" operators, $\hat{Q} \in {\cal L}^{(QH)}$. In
their full generality and without additional assumptions, they were
emphasized to exhibit peculiar and strongly counterintuitive, truly
``non-Hermitian" properties. In particular, once they were not
assumed to be Riesz operators or compact symmetrizable operators,
their spectrum was shown not to be necessarily real. Moreover, their
adjoints were shown non-quasi-Hermitian in general, $\hat{Q}^\dagger
\notin {\cal L}^{(QH)}$.

In the historical perspective it may be considered rather
unfortunate that more than thirty years later, Scholtz, Geyer and
Hahne \cite{Geyer} decided to choose, incidentally, {\em the same}
name for the well-behaved and very narrow compact-operator subset
${\cal L}^{(SGH)}$ of the Dieudonn\'{e}'s quasi-Hermitian set. This
is the reason why we use here the terminology inspired by Smilga
\cite{Smilga} and why we call the elements of the set ${\cal
L}^{(SGH)}$ cryptohermitian operators.

The most natural and physics-motivated extension of the
cryptohermitian class ${\cal L}^{(SGH)}$ is the Mostafazadeh's
\cite{Ali} set of the so called ``pseudo-Hermitian" operators ${\cal
L}^{(PH)}$ for which, roughly speaking, all of the invertible and
self-adjoint operators $\Theta$ compatible with Eq.~(\ref{dieudo})
(and denoted, therefore, by another symbol, say, ${\cal P}$) may
remain indefinite. A potentially equivalent name of ${\cal
PT}-$symmetric operators is usually reserved, by its proponents
\cite{DB,BG,BB}, to the narrower class of models where ${\cal P}$
coincides with the operator of parity and where it may be
interpreted as the indefinite metric in Krein space \cite{Bar}.

One of the most compact approaches to the applications of ${\cal
PT}-$symmetric operators  in quantum theory has been reviewed by
Carl Bender \cite{Carl}. He considers merely the intersection of the
cryptohermitian and ${\cal PT}-$symmetric classes of operators of
potential observables while picking up a unique metric in the
product form $\Theta^{(CB)}={\cal PC}$. The key merit of this
approach is that the factor ${\cal C}$ with the property ${\cal
C}^2=I$ may be interpreted as a charge of the system. The
alternative name for ${\cal C}$ as proposed in \cite{pseudo} (viz.,
``quasiparity") has not been given the meaning of an observable and
found much less users in the literature, therefore~\cite{Geza}.

\section*{Acknowledgement}

Work supported by the GA\v{C}R grant Nr. P203/11/1433, by the
M\v{S}MT ``Doppler Institute" project Nr. LC06002 and by the
Institutional Research Plan AV0Z10480505.


\newpage

\begin{thebibliography}{00}

\bibitem{Hoo}
J. Hilgevoord, Am. J. Phys. 70 (2002) 301.

\bibitem{discrsqw}
M. Znojil,
Int. J. Theor. Phys. 50 (2011) 1052.

\bibitem{amend}
M. Znojil,
J. Phys. A: Math. Gen. 30 (1997) 8771.

\bibitem{Actonbe}
for a characteristic sample of usage in molecular physics and
chemistry cf., e.g.,
 the MOLPRO software
package  http://www.molpro.net/, 
maintained by
 H.-J. Werner and P. J. Knowles.

\bibitem{AS}
M. Abramowitz and I. A. Stegun, Handbook of Mathematical Functions
(Dover, New York, 1970).
%
%


\bibitem{uNIST}
Frank W. J. Olver,
Daniel W. Lozier,
Ronald F. Boisvert,
Charles W. Clark, Eds.,
NIST Handbook of Mathematical Functions (Cambridge University Press,
Cambridge, 2010).

%
%


\bibitem{Maple}
B. W. Char et al, Maple V Language Reference Manual (Springer, New
York, 1993).

\bibitem{SIGMA}
M. Znojil,
SYMMETRY, INTEGRABILITY and GEOMETRY: METHODS and APPLICATIONS 
5 (2009)  001.


\bibitem{Dieudonne}
 J. Dieudonne, 
  Proc. Int. Symp.
            Lin. Spaces (Pergamon, Oxford, 1961), p. 115.




\bibitem{Geyer}
F. G. Scholtz, H. B. Geyer and F. J. W. Hahne, Ann. Phys. (NY) 213
(1992)
 74.


\bibitem{Carl}
C. M. Bender, Rep. Prog. Phys. 70 (2007) 947;


A. Mostafazadeh, 
Int. J. Geom. Meth. Mod. Phys. 7
(2010) 1191. 


\bibitem{Acton}
J. Stoer and R. Bulirsch. Introduction to Numerical Analysis
(Springer-Verlag, New York, 1980).



\bibitem{beNIST}
For tables of Gauss-Hermite abscissae and weights up to order n = 32
see
http://www.efunda.com/math/num\_integration/findgausshermite.cfm.



\bibitem{NIST}
NIST Digital Library of Mathematical Functions:
%
%
%
http://dlmf.nist.gov/


%


\bibitem{Smilga}
 A. V. Smilga, J. Phys. A: Math. Theor. 41 (2008) 244026.
%

\bibitem{Ali}
A. Mostafazadeh, 
J. Math. Phys. 43 (2002) 205, 2814  and 3944.

\bibitem{DB}
D. Bessis, private communication, 1992;

J. Zinn-Justin, private communication, 2010.

\bibitem{BG}
V. Buslaev and V. Grecchi, J. Phys. A: Math. Gen. 26 (1993) 5541.


\bibitem{BB}
C. M. Bender and S. Boettcher, Phys. Rev. Lett. 80 (1998) 5243.

\bibitem{Bar}
H. Langer and Ch. Tretter, Czechosl. J. Phys. 54 (2004) 1113.

\bibitem{pseudo}
M. Znojil, 
  arXiv:math-ph/0104012.

\bibitem{Geza}
G. L\'{e}vai, Czechosl. J. Phys. 54 (2004) 77;

A. Sinha and P. Roy, Czechosl. J. Phys. 54 (2004) 129;

O. Mustafa, Int. J. Theor. Phys. 47 (2008) 1300.



\end{thebibliography}
\end{document}